\documentclass[aps,prl,twocolumn,showpacs,superscriptaddress,groupedaddress]{revtex4}  
\usepackage{graphicx}  
\usepackage{dcolumn}   
\usepackage{bm}        
\usepackage{amssymb}   

\begin{document}


\author{Ofer Kimchi}
\affiliation{Department of Physics, Princeton University, Princeton, NJ, 08544}
\author{Sarah L. Veatch}
\affiliation{Department of Biophysics, University of Michigan, Ann Arbor, MI. 48109}
\author{Benjamin B. Machta}
\email{bmachta@princeton.edu}
\affiliation{Lewis Sigler Institute, Princeton University, Princeton, NJ, 08544}
\affiliation{Department of Physics, Princeton University, Princeton, NJ, 08544}

\title{Allosteric Regulation by a Critical Membrane}

\begin{abstract}


Many of the processes that underly neural computation are carried out by ion channels embedded in the plasma membrane, a two-dimensional liquid that surrounds all cells.  Recent experiments have demonstrated that this membrane is poised close to a liquid-liquid critical point in the Ising universality class.  Here we use both exact and stochastic techniques on the lattice Ising model to explore the ramifications of proximity to criticality for proteins that are allosterically coupled to Ising composition modes.  Owing to diverging generalized susceptibilities, such a protein's activity becomes strongly influenced by perturbations that influence the two relevant parameters of the critical point, especially the critical temperature.  In addition, the protein's kinetics acquire a range of time scales from its surrounding membrane, naturally leading to non-Markovian dynamics.

\end{abstract}
\maketitle


Every cell is surrounded by the plasma membrane, a two dimensional (2D) liquid composed of lipids and embedded proteins.  
In addition to separating the cell from its surroundings, the plasma membrane is home to diverse functional processes.  In neurons, membrane-bound ion channels sense chemical and electrical signals and control conductance to specific ions, leading to the complex dynamics that underly neural function.  
Recent progress suggests a role for the membrane itself in regulating these processes.  The membrane is heterogeneous, with liquid structures often termed `rafts' at length scales of $10-100$ nm, much larger than the $1$ nm scale of individual lipids~\cite{LipidRaftsProteins,LipidRaftsIonChannels}.  Experiments have suggested a physical mechanism underlying these structures.  Vesicles isolated from a mammalian cell line have membranes tuned close to a liquid-liquid miscibility critical point~\cite{Veatch08}.  When cooled below their critical temperature, $T_c$, these vesicles macroscopically phase separate into two liquid phases termed liquid ordered ($l_o$) and liquid disordered ($l_d$) which differ in the partitioning of lipids, proteins and a fluorescent dye~\cite{BaumgartHSHHBW07}.  

Previously we have argued that proximity to this critical point is likely to underly much of the `raft' heterogeneity seen in diverse membrane systems~\cite{Machta11} with embedded proteins subject to long-range critical Casimir forces~\cite{Machta12}.   More recently we have shown that n-alcohol anesthetics take membrane derived vesicles away from criticality by lowering $T_c$~\cite{Gray13}. Despite structural diversity, anesthetics are known to exert similar effects on diverse ion channels~\cite{Franks94} leading us to speculate that these effects might arise because anesthetics mimic or interfere with native regulation of channels by their surrounding membrane.  In support of this, we found that conditions that reverse anesthetic effects on ion channels and organisms also raise critical temperatures in vesicles~\cite{Machta16}.

\begin{figure}
	\includegraphics[width=\columnwidth]{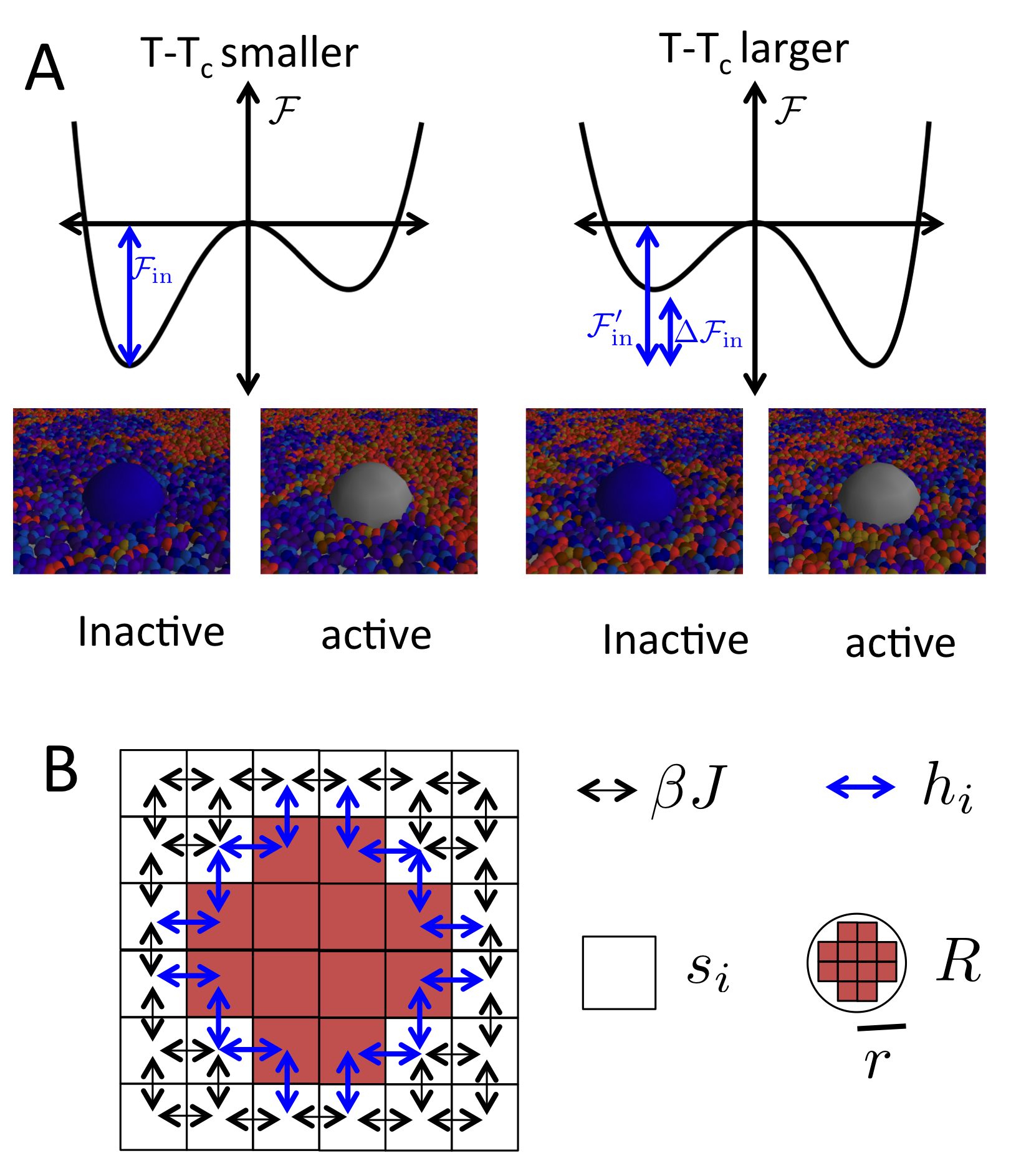} 
	\caption{\label{fig:model} (A) Schematic representation of our model.  We consider a protein embedded in a nearly critical membrane with which it interacts through distinct boundary interactions in distinct functional states.  When the parameters of the membrane change, the free energies of different protein states changes, altering function. 
(B) We probe these effects using a lattice Ising model in which a single protein is modeled as a group of sites which transition together, while the remaining membrane is composed of Ising spins that can take values $s_i= \pm 1$}
\end{figure}

Here we explore consequences of thermodynamic criticality for a membrane-bound protein whose internal state is coupled to the state of its surrounding membrane.  
While most ion channels are broadly classified as ligand gated or voltage gated, many are also sensitive to a wide range of modulators including calcium levels, pH, lipids, and temperature \cite{Kinnunen91}.
Ion channel function can also depend on the 2D solvent properties of the membrane in which they are embedded, both in reconstituted \cite{Bristow87,Rankin97} and \textit{in vivo} assays~\cite{Sooksawate01b,Allen07}. In this manuscript we develop a simple model for a protein regulated by its surrounding membrane.  We show that when the membrane is held close to a critical point this type of coupling leads to a qualitatively distinct regime including strong responses to perturbations which influence $T_c$, a hierarchy of time scales, and non-Markovian dynamics.  As we show, these effects are critical phenomena, arising from the diverging generalized susceptibilities and correlation times that emerge near the critical point.  
 
Throughout the manuscript we consider a system consisting of membrane degrees of freedom $\{s\}$ and a single hypothetical ion channel (Fig. 1A) with state $R$ and Hamiltonian
\begin{equation}
\mathcal{H}_{tot}(\{s\},R)=\mathcal{H}_{mem}(\{s\})+E_R+\mathcal{H}_{int}(\{s\}_\partial,R)
\end{equation}
where $\mathcal{H}_{mem}$ describes interactions of membrane lipids with one another, $E_R$ measures the (free) energy of state $R$ without considering membrane interactions, and $\mathcal{H}_{int}$ describes the interaction between the protein in state $R$ and the components that it contacts at its boundary, $s \in \partial$.  In equilibrium, the system consisting of protein and membrane will be in a given state with probability determined by an appropriate Boltzmann distribution, $P(\{s\},R)=e^{-\beta \mathcal{H}_{tot}(\{s\},R)}/Z$.  We can integrate over membrane degrees of freedom to isolate the protein whose internal states are occupied according to $P(R)=\exp(-\beta \mathcal{F}_R)/Z$, where
\begin{equation}
\exp(-\beta \mathcal{F}_R)=\sum_{\{s\}} \exp{(-\beta \mathcal{H}_{tot}(\{s\},R))}
\end{equation}
defines $\mathcal{F}_R$, the free energy of  state $R$. 

As in previous work, we model the membrane as a square-lattice of spins $s_i=\pm 1$ with the usual Ising Hamiltonian $\mathcal{H}_{mem}=-J\sum_{\langle i,j \rangle } s_i s_j $, where the sum runs over all nearest neighbor pairs neither of which are contained in the protein.  In our model, up/down spins roughly correspond to more $l_o / l_d$ partitioning components, with a lattice constant which corresponds to $l \sim1-2 \text{  nm}$ of membrane. This scale corresponds both to the approximate size
of a lipid of $\sim .8$ nm$^2$\cite{lipidSize,lipidSize2} and to measurements of the correlation length near the critical point of cell-derived vesicles that suggest an analogy to a lattice model with $l \sim 2$ nm ~\cite{Veatch08}.  Our model also contains a protein whose internal state can take two values, $R=\pm 1$ which could correspond to open/closed, and a chemical potential (perhaps modulated by a ligand) that contributes a term in the Hamiltonian $E_{R} = -\mu R$.  Crucially we also include an interaction between membrane and protein of the form $\mathcal{H}_{int} =\sum_{i} h_i(R) s_i$ where this sum is over spins which border the protein, whose boundary conditions (BCs) in state $R$ are determined by  $h_i(R)$ (See Fig. 1B).  These interactions between our protein's state $R$ and surrounding spins mimic a coupling between a protein's functional state and the membrane's state as seen experimentally~\cite{Bristow87,Sooksawate01b,Allen07}. They could arise from hydrophobic mismatch~\cite{Andersen07,Soubias08},
distinct lateral pressure profiles in the different phases~\cite{Gruner91}, 
or specific interactions with particular components that partition strongly into distinct phases~\cite{Levitan14}.

We consider three types of BCs in this manuscript: \textit{fixed} BCs in which $h_i(R)= \pm J_R$, \textit{free} BCs in which $h_i(R)=0$ and \textit{Janus} BCs in which $h_i(R) = +J_R$ on one side of the protein and $h_i(R) = -J_R$ on the other side.

We first wanted to investigate how a change in the membrane's properties might influence average channel activity, as would be measured by a whole cell recording or other technique probing the response of many channels together.  In the context of our model this means examining the dependance of $P(R)$ on the details of $\mathcal{H}_{mem}$.  
 As this is a static property of our equilibrium system we can make use of a remarkable algorithm developed for spin glasses that uses Pfaffian elimination to exactly calculate partition functions to user specified precision on lattices with arbitrary nearest neighbor couplings $J_{ij}$ in zero field~\cite{Creighton13}.  We implement our Hamiltonian by setting all interactions $J_{ij}$ between spins to $\beta J$, interactions between lattice sites internal to the protein to a large negative value (here  $-10$) and interactions between lattice sites contained in the protein and spins to $h_i(R)$ \footnote{This procedure more precisely calculates $\beta \mathcal{F}+\log 2$ since it integrates over both the correct boundary conditions and those where every $h_i$ is flipped.} (see Fig. 1B). 

Changes to $\mathcal{H}_{mem}$ will lead to a change in a state's free energy $\Delta \mathcal{F}_R$ with resulting changes in $P(R)$.  In this two-state example, only the difference in free energy changes, $\Delta  \mathcal{F}_{diff}=\Delta \mathcal{F}_+ -\Delta \mathcal{F}_-$ will influence $P(R=\pm1)$ through: 
\begin{equation}
\frac{P(R=+1)}{P(R=-1)} \rightarrow  \frac{P'(R=+1)}{P'(R=-1)} \exp \left(-\beta \Delta \mathcal{F}_{diff}\right)
\end{equation}
Thus $\exp (-\beta \Delta \mathcal{F}_{diff})$ measures the degree to which a change in $\mathcal{H}_{mem}$ potentiates the $+$ state (see Fig. 2A).

We estimate the potentiation of function that accompanies a $1\%$ change in $T_c$, comparable both to the natural variation seen in cell derived vesicles~\cite{Veatch08}, and to the magnitude of the effect seen with physiologically relevant concentrations of anesthetics~\cite{Gray13}.  We examine BCs in which the $R=-1$ state has fixed BCs ($h_i=+\infty$, here $10$ for numerical purposes) while the $R=+1$ state has either free BCs ($h_i=0$) or Janus BCs. The results are shown in $\text{Fig. 2B-D}$.  In $\text{Fig. 2B}$ this potentiation is shown for proteins in which the $R=+1$ has free BCs, at $T=1.05 T_c$ for proteins with a range of radii.  Larger proteins see more pronounced potentiation due to the increased surface area of interaction. In $\text{Fig. 2C}$ the potentiation is shown for the same proteins over a range of temperatures.  Away from the critical point, this change has only a small effect on the state occupancy -- the potentiation is small.  However, near the critical point the potentiation becomes much larger, even for the modest $1\%$ change in $T_c$ explored here.  This change in $T_c$ is comparable to that seen with addition of clinically relevant concentrations of anesthetic~\cite{Gray13,Machta16}. For the largest proteins sizes explored here, our observed potentiation is similar to that observed for the GABA$_A$ channel (see Fig. 2 of Ref~\cite{Franks94}) whose radius is $\sim 6$ nm.  While we don't know details of GABA$_A$'s interaction with it's surrounding membrane, our results suggest that anesthetic effects on membrane $T_c$ could in principle lead to observed changes in channel function.

\begin{figure}
	\includegraphics[width=\columnwidth]{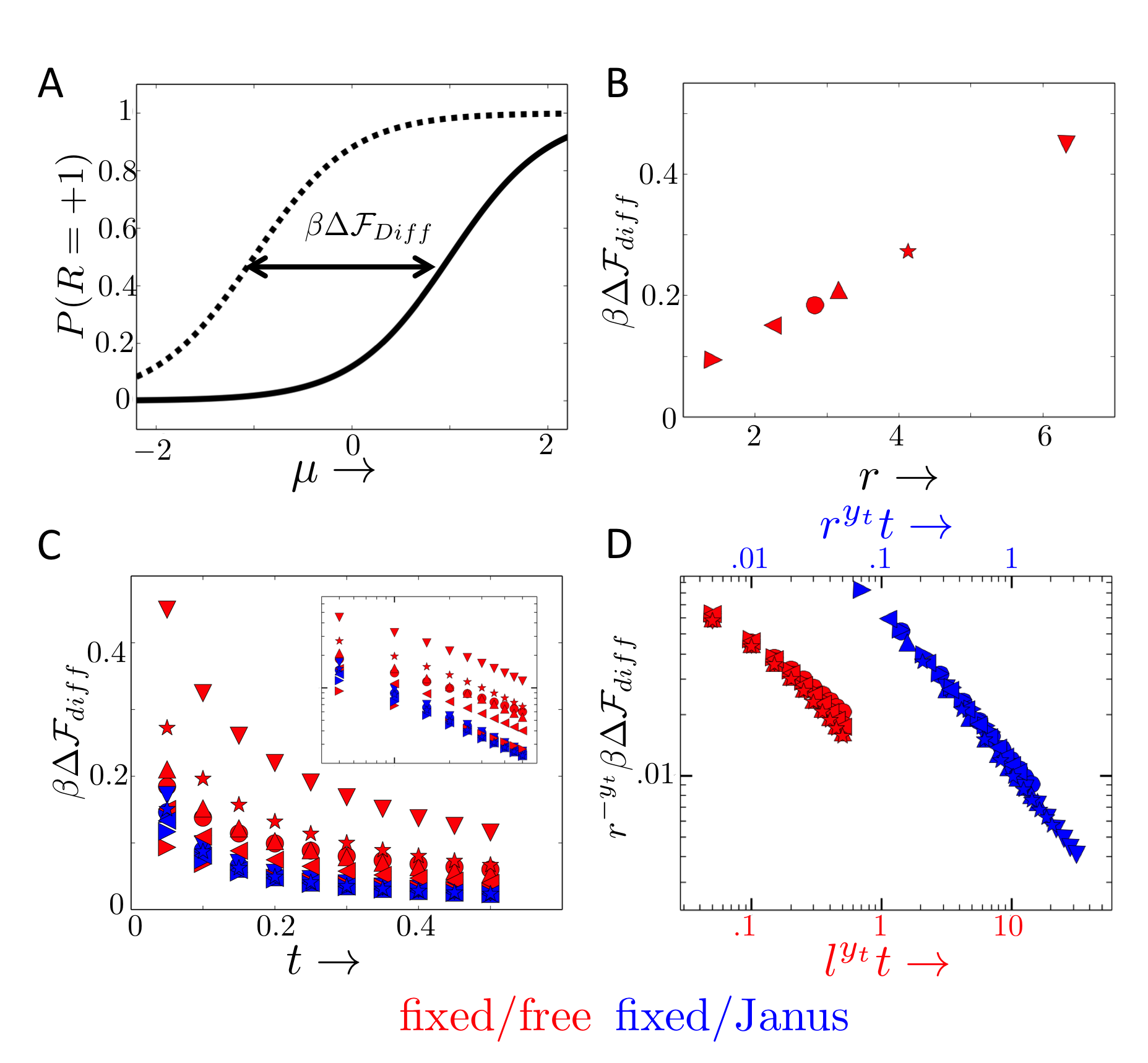}
	\caption{\label{fig:2} Changes in membrane properties lead to changes in the conformational equilibrium of a membrane-bound protein.  (A) In this example $P(R=+1)$ is potentiated by a change in the critical temperature of the membrane.  As $\mu$ is varied (for example through addition of ligand) the protein transitions from the off ($R=-1$) to on ($R=+1$) state.  After a perturbation has been applied that changes $T_c$, the free energy difference between the $R=\pm1$ states has changed by $\beta \Delta \mathcal{F}_{diff}$, shifting the curve of $P(R=+1)$ vs $\mu$ to the left (from the black line to the black dashed line). (B) $\beta \Delta \mathcal{F}_{diff}$ is plotted for a perturbation in which $T_c$ is lowered by $1\%$ for ion channels of different sizes, in all cases from $T=1.05 T_c$ and with fixed boundary conditions when $R=-1$ and free boundary conditions when $R=+1$.  $\beta \Delta \mathcal{F}_{diff}$ is larger for larger inclusions, and is $\sim 0.5$ for the largest inclusions examined, comparable to the GABA$_A$ channel which is $\sim6$ nm across and which is potentiated by $\sim 50\%$ by analogous treatments~\cite{Franks94}. (C) $\beta \Delta \mathcal{F}_{diff}$ is plotted vs $t=T-T_c/T_c$ for different sized inclusions (symbols as in B) for the same fixed-free BCs (red) and for BCs in which $R=+1$ has Janus BCs (blue).  In each case, the magnitude of $\beta \Delta \mathcal{F}_{diff}$ becomes much larger closer to the critical point. (D) We verify through the scaling collapse discussed in the text that this effect can be understood as a critical phenomenon.}
\end{figure}

We next wanted to verify that our effect could be understood as a critical phenomenon.  According to scaling, the free energy associated with the insertion of an inclusion 
should have a singular contribution that depends on inclusion size, BCs and the distance to the critical point:
\begin{equation}
\mathcal{F}^s_{R} = \mathcal{U}(r t^{1/y_t},  h/t^{y_h/y_t},...)
\end{equation}
where $y_{t}=\nu=1$ and $y_h =\gamma = 15/8$ are the scaling dimension of the fields $t$ and $h$, $r$ is the radius of the inclusion and $\mathcal{U}$ is a universal function.  Taking a derivative with respect to t, we expect that 
\begin{equation}
\frac{\partial \mathcal{F}^s_{diff}}{\partial t} = r^{y_t} \mathcal{U}(r t^{1/y_t}, h/t^{y_h/y_t},...)
\end{equation}

We verify that our measured potentiation can indeed be collapsed in this manner in Fig. 2D for the fixed/Janus BCs.  However, for the fixed/free particles we find an additional contribution that arises more directly from the boundary.  This term is proportional to the length of the boundary and should scale as 
\begin{equation}
\mathcal{F}^s_{R,\partial} = r/l \text{ } \mathcal{U}(l t^{1/y_t},  h/t^{y_h/y_t},...)
\end{equation}
leading to a similar collapse but with $l^{y_t} t$ on the x-axis (see supplement for discussion). While scaling suggests that coupling to changes in $h$ would diverge more strongly near criticality, membranes are at fixed magnetization (composition) in which there is no divergence.


We next looked at the role a critical membrane could play in shaping the kinetics of ion channel function as would be measured in single channel voltage clamp experiments. To do this we employ the local Kawasaki algorithm \cite{Kawasaki72} as we have done previously \cite{Machta11} which implements diffusive `model B' dynamics \cite{HohenbergHalperin}. Using these dynamics a sweep (in which every spin is proposed to exchange with its neighbors once) corresponds to roughly a microsecond of real time since each lattice site corresponds to the size of a lipid and lipids diffuse at a rate of $\sim 1\ \mu \text{m}^2$/s \cite{Machta11}\footnote{One caveat is that cells are likely in the regime where hydrodynamic relaxation \cite{HonerkampSmithHydrophobic} competes with diffusive dynamics.}.   Our Kawasaki dynamics forbid exchange with spins in the protein, but every sweep the protein changes state with the Metropolis probability to ensure detailed balance. 

While our protein attempts to change its state $R$ during every sweep, its apparent kinetics are much slower.  Representative traces of $R(\tau)$ are shown in $\text{Fig. 3A}$ after convolution with a Gaussian of width $\sigma=10^3$ sweeps
~\footnote{Because patch clamp recordings measure times on the order of 50-100 $\mu s$ \cite{patchClamp,Kasianowicz96,Sakmann09}, we plot $R(\tau)$ after convolution with the Gaussian $k(\tau) = e^{-\tau^2 / \sigma^2}$ where $\sigma=10^3$ (see supplement for bare curves and a discussion). Thus, while our model has only two states, the recording shows short spikes reaching levels in between the two, a feature common to real recordings that presumably has a similar explanation \cite{patchClamp}.}.  These traces share qualitative features with real ion channels spanning many time scales~\cite{Mortensen07}: `flickers' shorter than $\sigma$, single `openings'  (which often contain many short flickers), and `bursts' in which a series of openings occur close to each other. The time-scale of these features increases as the critical point is approached ($\text{Fig. 3A}$). Similar changes in the time-scale of ion channel dynamics arising from the addition of some anesthetics have been observed experimentally~\cite{Wachtel95}. These could result from perturbations to the proximity of the membrane to criticality, though they are typically interpreted as arising from direct binding interactions between anesthetic and channel.


We quantify these findings by looking at the time auto-correlation of the simulated protein, $\chi(\tau) =\left< R(0) R(\tau) \right>$, for a range of temperatures ($\text{Fig. 3B}$), demonstrating that correlations in state persist much longer as the critical point is approached.  We also perform simulations for proteins of different radii all at $T=1.05T_c$ ($\text{Fig. 3C}$), demonstrating that larger proteins have slower kinetics due to the increased area for interaction with surrounding membrane.  These curves quantify the many time-scales that can be seen qualitatively in $\text{Fig. 3A}$; they are not closely approximated by single exponentials.  Although the system as a whole is Markovian, when considered in isolation the protein displays non-Markovian dynamics, implying that some memory of its history is stored in the membrane degrees of freedom surrounding it. The non-Markovianity of several real ion channels and other membrane-embedded proteins has been well characterized and various internal states of these proteins have been hypothesized \cite{Yamashita05,MillerCholesterol}; however, our results suggest that in some cases the history of the protein's state may be stored in the membrane rather than in internal states of the protein itself.

We expect that the long time-scales seen in $\text{Fig. 3A}$ are inherited from the surrounding membrane.  Near the Ising critical point the correlation length diverges as ${\xi \propto t^{-\nu}}$, and the correlation time diverges like $\tau_{\text{cor}} \propto \xi^z$ where a \textit{dynamic} exponent is $z=4-\eta= 3.75$ for the Kawasaki dynamics employed here \cite{HohenbergHalperin}. If the curves of $R(t)$ reflect properties of these slow critical fluctuations, then we might expect that autocorrelation functions in systems with different radii and correlation lengths $\chi(\tau,r, \xi)$ might take a universal form:
\begin{equation}
\label{eq:UniversalTime}
\chi(\tau, r,\xi)=r^\lambda \mathcal{U} \left(r/\xi, \tau/\xi^z \right)  
\end{equation}
with $\lambda=2y_{\text{b}}=1$ where $y_{\text{b}}$ is the scaling dimension of the boundary operator which is expected to be $1/2$ from conformal field theory arguments (See~\cite{Cardy04} and supplement for a brief discussion).  To check this, we performed simulations for the same proteins shown in Fig. 3C but now at different temperatures chosen so that $r/\xi$ is fixed.   While the bare curves decay over time-scales that differ by a factor of $10^2$, when we plot $\chi(\tau)/r$ vs $\tau/\xi^z$,  we see collapse onto a single curve within stochastic error ($\text{Fig. 3D}$) as predicted by equation \ref{eq:UniversalTime}.

\begin{figure}
	\includegraphics[width=\columnwidth]{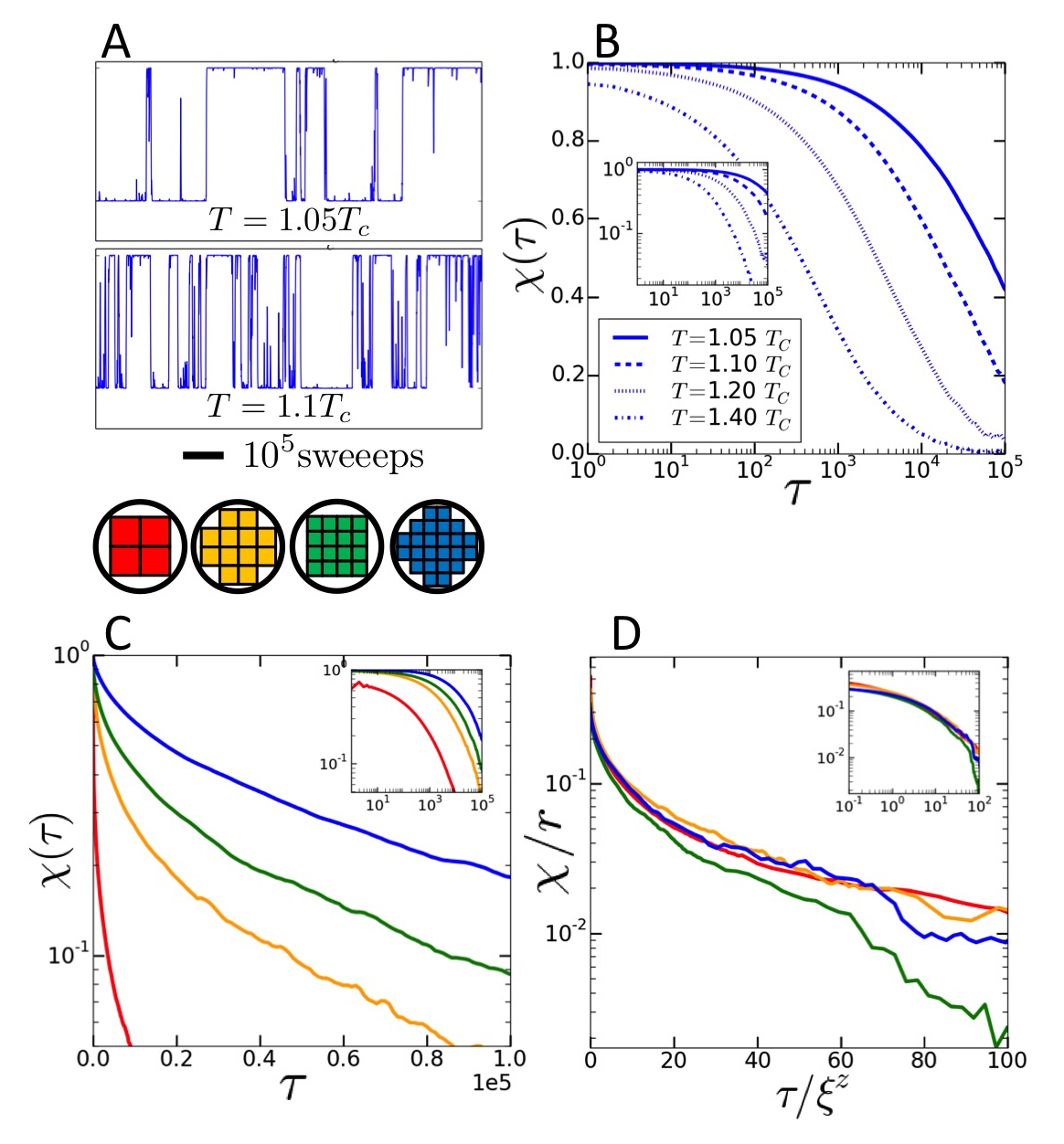}
	\caption{\label{fig:dynamics} (A) Simulated ion channel dynamics, $R(\tau)$, have prominent features at many distinct time-scales, qualitatively resembling traces from real single-channel recordings~\cite{Mortensen07}. Near $T_c$, protein state changes occur on a timescale orders of magnitude larger than that of attempted state changes which happen once per sweep. (B) The autocorrelation function of $\chi(\tau)$ is shown at different temperatures, with proximity to criticality leading to longer lived correlations. (C) $\chi(\tau)$ also depends on channel radius $r$ at a single temperature, $T= 1.05 \ T_c$ (colors as above). 
(D) These long time scales are critical phenomena. When we plot $\chi(\tau)/r$ vs $\tau/\xi^z$ for $r$ and $\xi$ values chosen so that $r/\xi$ is constant, these curves collapse onto a single universal function as predicted by scaling. }
\end{figure}

\textit{Conclusion-}  Regulation by the nearly critical membrane might be very widespread both in ion channels and for other membrane-bound proteins.  There are hints that the membrane may be playing an important role in this localization~\cite{Li07} and in some cases in direct regulation~\cite{Allen07} of channels.  Our model could account for the sensitivity of many diverse ion channels to chemically diverse anesthetics~\cite{Franks94} which we have demonstrated lower the critical temperature of cell-derived vesicles by $\sim4$K, just over $1\%$~\cite{Gray13}. This explanation is appealing in that it might also explain why cholesterol depletion, which changes the membrane's `magnetization'~\cite{Zhao13} also effects many anesthetic-sensitive channels~\cite{Sooksawate01b}. Our results suggest that such channels should have their partitioning into small domains modulated by addition of ligand, a prediction that could be tested using super-resolution techniques.

While most previous studies, both by ourselves \cite{Machta11,Machta12} and others~\cite{Shachi16} have focused on the role of membrane thermodynamics in localizing proteins into domains, this work suggests the same domains could couple more directly to function.  We predict that when a receptor binds ligand it will change its preference for its surrounding lipid environment, leading to different interaction partners and an imprint on the state of the local membrane that will persist even after the ligand has dispersed. We have highlighted a specific impact of thermodynamic criticality on a single ion channel, and further work will clarify how this mechanism contributes to neural function. 



\section{Acknowledgements}
We thank Bill Bialek, Colin Clement, Jim Sethna and Ned Wingreen for useful discussions.  This work was supported by NIH R01 GM110052 (SLV) and a Lewis-Sigler Fellowship (BBM).  BBM thanks CK Thomas and AA Middleton for making their code available on their website (https://aamiddle.expressions.syr.edu/).

\clearpage

\section{Supplemental Information}

In supplemental information we first provide a more detailed argument for the form of the scaling function needed to capture fixed/free BCs used to make Fig. 2D of the main text.  We then discuss the scaling form of the dynamic correlation function $\chi( \tau, r ,\xi)$.  Finally, we discuss subtleties in our gaussian blurring procedure used to make Fig. 3A in the main text.

\subsection{Scaling of $\partial \mathcal{F} / \partial J$ for free BCs}

In a suitable continuum limit we can write the Hamiltonian conditioned on particular boundaries as:

\begin{equation}
H=J \int_{|\vec{x}|>r} d\vec{x} \epsilon (\vec{x})
\end{equation}

where $\epsilon(\vec{x})$, the energy operator, is the continuum version of $\epsilon_{ij}=s_{i,j}s_{i+1,j}$.  Both here and on the lattice, we can write the free energy change associated with a change in Hamiltonian parameters as:

\begin{equation}
\frac{\partial \mathcal{F}}{\partial J}=\left<\int d\vec{x} \epsilon (\vec{x}) \right>
\end{equation} 

Here we use conformal field theory techniques to exactly calculate the field $\epsilon(\vec{r})$ that surrounds inclusions with our fixed and free BCs exactly at the critical point and in the continuum.  Although we are unable to carry out this calculation away from $T_c$, we can use scaling to find the form of the relevant function. 

At the critical point we make use of exact methods that we and others have previously used to calculate Casimir forces between inclusions with particular BCs.  Here we take a different limit, wherein one of the inclusions is infinitesimal and serves only as a `test particle' for measuring the strength of the the other inclusion's induced field $\epsilon(\vec{d})$, which is a function of its BCs.

To calibrate our test particle we first note that $\epsilon(\vec{d})$ is a primary scaling field with scaling dimension $1$ so that:
 
\begin{equation}
\left<\epsilon(0)\epsilon(\vec{d})\right> \sim \frac{1}{|\vec{d}|^2}
\end{equation}  

An inclusion with Free BCs in the Ising model will not couple to the order parameter, and so its leading contributions arise from its coupling to the energy field $\epsilon(\vec{x})$.  When two inclusions of radius $r$ are placed a large distance $d$ apart, their interactions are closely approximated as arising from point like insertions of field so that their interaction energy goes like:

\begin{equation}
U_{Fr-Fr}(d,r,r) \sim h_\epsilon^2(r) \left<\epsilon(\vec{0}) \epsilon(\vec{d}) \right> + \text{higher order in d}
\end{equation}  
This in conjunction with the exact results allows us to infer the coupling $h_\epsilon(r)\sim r$.

Next we use the same exact results this time with one protein of radius $r$ and one of infinitesimal radius $a_0$.  At the critical point, conformal invariance allows implies that their interaction energy can be written as a function of just one number, $x=\frac{(d+2a_0)(d+2r)}{ra_0}$.  As $a_0$ becomes small, $x$ becomes large, even when $d<<r$, allowing for a measurement of $\epsilon(r)$ everywhere in space including closeby to our inclusion of interest.  In this limit, the scaling form is sufficient, and we find:
\begin{equation}
\left<\epsilon(\vec{d})\right> = \frac{r}{|\vec{d}|(|\vec{d}|+2r)}
\end{equation}
Where the expectation values are conditioned on the presence of an inclusion with Free BCs of radius $r$ at the origin, and where $d$ measures the distance from the surface of the protein.  We are interested in the integral over space of $\left<\epsilon(\vec{d})\right>$ which diverges at both large and short distances.  The short distance divergence we will cut off by starting our integration at a cutoff motivated by a lattice at $l_0$.  Since we are primarily interested in the nearly critical, rather than critical regime, we assume that the off-critical energy field takes the following form, which is certainly true in a regime where $\xi>r>l$:
\begin{equation}
\left<\epsilon(\vec{d})\right> = \frac{r}{|\vec{d}|(|\vec{d}|+2r)} \times \mathcal{U}(d/\xi, r/\xi, l/\xi,...)
\end{equation}
where the universal function presumably depends primarily on its first argument (the latter will henceforth be omitted).  Though we do not know the exact form of $\mathcal{U}(d/\xi)$ we expect that it is close to $1$ when $d/\xi <<1 $ and that it goes to 0 when $d/\xi >>1$. Our desired integral can now be written
\begin{equation}
\frac{\partial \mathcal{F}}{\partial J}=\left<\int d\vec{d} \epsilon (\vec{d}) \right>= \int^\infty_{l_0} 2 \pi (r+|\vec{d}|) d|\vec{d}| \frac{r}{d(d+2r)} \mathcal{U}(d/\xi)
\end{equation} 
Now we split this into two terms determined by whether their contribution diverges as $l \rightarrow 0$ or only as $\xi \rightarrow \infty$. 
\begin{equation}
\int^\infty_{l_0} \pi d|\vec{d}| \frac{r}{d} \mathcal{U}(d/\xi) 
\end{equation}
and
\begin{equation}
\int^\infty_{l_0} \pi d|\vec{d}| \frac{r}{(|\vec{d}|+2r)} \mathcal{U}(d/\xi) 
\end{equation}

 In the first we change coordinates to $x=d/l_0$

\begin{equation}
\pi r \int^\infty_{1} dx \frac{\mathcal{U}\left( (l/\xi)\ x\right)}{x} 
\end{equation}
which is of the form $r\ \mathcal{U}_1 (l/\xi)$ for a (different) universal function $\mathcal{U}$.  For the second term, we can take $l \rightarrow 0$ and instead change variables to $x=d/r$ yielding
\begin{equation}
\pi r \int^\infty_{0} dx \frac{\mathcal{U}\left( (r/\xi)\ x\right)}{(x+2)} 
\end{equation}
which is of the form $r\ \mathcal{U}_2 (r/\xi)$. 

From the above we can write
\begin{equation}
\frac{\partial \mathcal{F}}{\partial J}=r\ \mathcal{U}_1 (l/\xi) + r\ \mathcal{U}_2 (r/\xi)
\end{equation} 

As both terms diverge as the arguments of $\mathcal{U}_{1/2}$ become large, we generally expect the first to dominate when except when it is 0 for some reason.  This is the case for the fixed/free BCs as can be seen in Fig. 2D of the main text.  This calculation does not easily extend to the Janus/fixed case, where (for example) the field $\epsilon(\vec{d})$ presumably has angular dependance.  However, since the boundary of the janus BCs look locally similar to the boundary of the fixed BCs, we might expect that the term with $l/\xi$ dependance cancels, leaving only a term with $r/\xi$ dependance, as is seen in Fig.2D of the main paper.

\subsection{Scaling of dynamics with radius}

We wish to argue for the scaling form of ${\chi(\tau,r,\xi)=\left<R(0)R(\tau)\right>}$ defined in the paper.  First consider a simpler version where a site magnetic field adds a term in the Hamiltonian $-h s_0 R$.  We assume that at a rate $k$, $R$ switches or not so as to satisfy detailed balance so that:
\begin{equation}
<R(0)R(\tau)>=\mathcal{U}(h, k \tau, \tau/\xi^z)
\end{equation}
If we assume $k$ is very fast we can ignore the $k$ dependance.  Furthermore, $h$ dependance must be even, so that when $h$ is small we have:
\begin{equation}
<R(0)R(\tau)>=h^2\mathcal{U}(\tau/\xi^z)+\mathcal{O}(h^4)
\end{equation} 
where, in fact, this is just $\left<R(0)R(\tau) \right>=h^2 \left<s_0(0)s_0(\tau) \right>$.  

We would like to do a similar calculation here, but now we must replace $h$ with $h_{eff}(r, J_r)$.  We suspect that ours is best understood as a `boundary field', whose dimension is $y_{b}=1/2$~\cite{Cardy04}  so that $h_{eff}(r)\sim r^{1/2}J_r$ so that 
\begin{equation}
\chi(\tau,r,\xi)=\mathcal{U}(r^{y_b}J_r, k\ \tau, \tau/\xi^z)
\end{equation}
which, for small $r^{1/2}J_r$, small $k$ and fixed $J_r$ can be written as 
\begin{equation}
\chi(\tau,r,\xi)=r^{2y_b}\mathcal{U}(\tau/\xi^z)
\end{equation}
This analysis predicts that had we also scaled $J_r \sim r^{-y_b}$ that we could have found results that collapsed even for larger proteins for which our relatively large $J_r$s almost never flip.

\subsection{Gaussian Blurring}

\begin{figure}[tb]  
\centering	
			\includegraphics[scale=0.35]{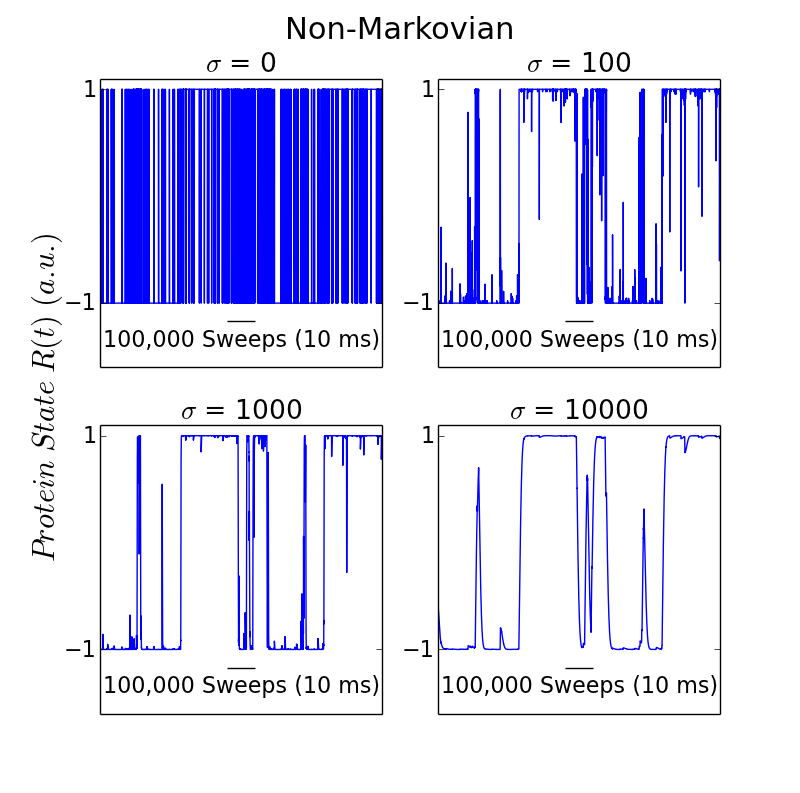}
			\includegraphics[scale=0.35]{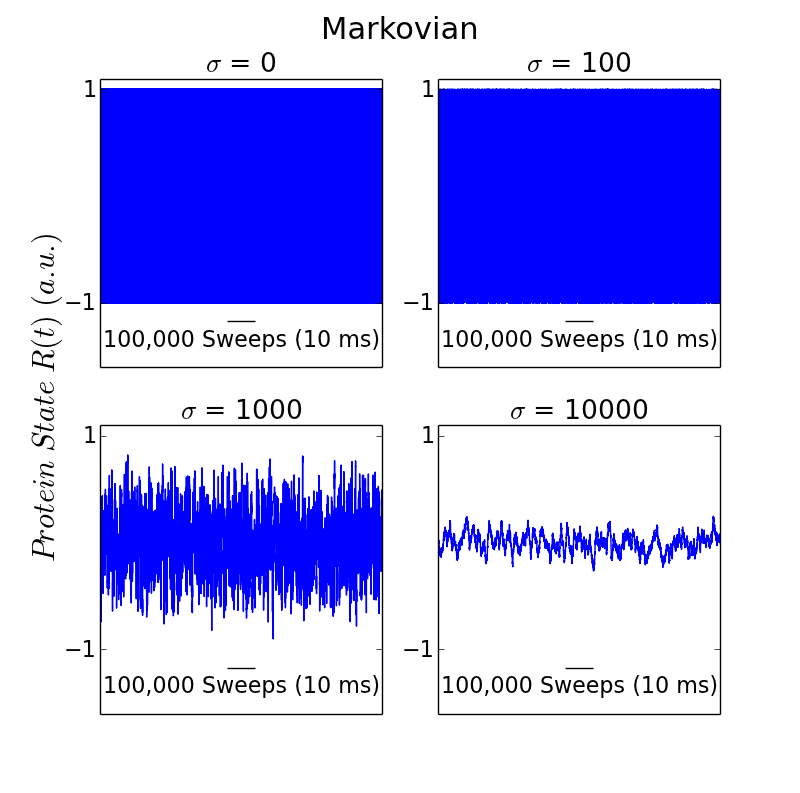}
\caption{\textbf{Results Are Robust to Resolution of Gaussian Blurring} Top: The simulated protein attempted to flip every sweep (corresponding to $0.1 \ \mu s$), and without any blurring, the simulated ion channel recording includes state changes on the corresponding timescale (top left). However, when the low temporal resolution of standard patch-clamp techniques is accounted for through convolution of the data with the Gaussian $k(t) = e^{-t^2 / \sigma^2}$ as in Fig. 3A, a timescale for protein dynamics on the order of tens of $ms$ can be seen. This result is not changed when the resolution of the Gaussian blurring, $\sigma$, is varied, even by orders of magnitude. Bottom: No such timescale emerges in the Markovian case. }
\label{fig:diffSigmas}
\end{figure}

Patch clamp recordings measure times on the order of 50-100 $\mu s$ \cite{patchClamp, Kasianowicz96, Sakmann09} while the dynamics of the ion channels we simulate here occur on timescales of 0.1 $\mu s$. Therefore, in order to simulate the comparitively low temporal resolution of patch clamp recordings, we plot the state of the channel as a function of time, $R(t)$, after convolution with the Gaussian $k(\tau) = e^{-\tau^2 / \sigma^2}$. In Fig. \ref{fig:diffSigmas} (top), we show the resulting simulated recordings for a range of $\sigma$s spanning several orders of magnitude.  Gaussian blurring leads to the emergence of a timescale of ion channel dynamics on the order of tens of $ms$, and this result is robust to changes in $\sigma$. This feature appears to be a non-Markovian phenomenon: a similar analysis performed on an ion channel simulated at high temperature far from criticality, for which the channel's dynamics are Markovian, does not lead to the emergence of such timescales (Fig. \ref{fig:diffSigmas}, bottom).

\begin{thebibliography}{10}

\bibitem{LipidRaftsProteins}
Simons K, Toomre D
\newblock (2000) Lipid rafts and signal transduction.
\newblock \emph{Nature Reviews Molecular Cell Biology} 1:31--39.

\bibitem{LipidRaftsIonChannels}
Dart C
\newblock (2010) Lipid microdomains and the regulation of ion channel function.
\newblock \emph{The Journal of Physiology} 588:3169--3178.

\bibitem{Veatch08}
Veatch SL, {et~al.}
\newblock (2008) Critical fluctuations in plasma membrane vesicles.
\newblock \emph{Acs Chemical Biology} 3:287--293.

\bibitem{BaumgartHSHHBW07}
Baumgart T, {et~al.}
\newblock (2007) Large-scale fluid/fluid phase separation of proteins and
  lipids in giant plasma membrane vesicles.
\newblock \emph{Proc. Natl. Acad. Sci.} 104:3165--3170.

\bibitem{Machta11}
Machta BB, Papanikolaou S, Sethna JP, Veatch SL
\newblock (2011) Minimal model of plasma membrane heterogeneity requires
  coupling cortical actin to criticality.
\newblock \emph{Biophys J} 100:1668--77.

\bibitem{Machta12}
Machta BB, Veatch SL, Sethna JP
\newblock (2012) Critical casimir forces in cellular membranes.
\newblock \emph{Phys. Rev. Lett.} 109:138101.

\bibitem{Gray13}
Gray E, Karslake J, Machta BB, Veatch SL
\newblock (2013) Liquid general anesthetics lower critical temperatures in
  plasma membrane vesicles.
\newblock \emph{Biophys J} 105:2751--9.

\bibitem{Franks94}
Franks NP, Lieb WR
\newblock (1994) Molecular and cellular mechanisms of general anaesthesia.
\newblock \emph{Nature} 367:607--14.

\bibitem{Machta16}
{Machta} BB, {et~al.}
\newblock (2016) {Stabilizing membrane domains antagonizes n-alcohol
  anesthesia}.
\newblock \emph{In press, Biophys J.}

\bibitem{Kinnunen91}
Kinnunen PK
\newblock (1991) On the principles of functional ordering in biological
  membranes.
\newblock \emph{Chemistry and Physics of Lipids} 57:375 -- 399.

\bibitem{Bristow87}
Bristow DR, Martin IL
\newblock (1987) Solubilisation of the gamma-aminobutyric acid/benzodiazepine
  receptor from rat cerebellum: optimal preservation of the modulatory
  responses by natural brain lipids.
\newblock \emph{J Neurochem} 49:1386--93.

\bibitem{Rankin97}
Rankin SE, Addona GH, Kloczewiak MA, Bugge B, Miller KW
\newblock (1997) The cholesterol dependence of activation and fast
  desensitization of the nicotinic acetylcholine receptor.
\newblock \emph{Biophysical Journal} 73:2446--2455.

\bibitem{Sooksawate01b}
Sooksawate T, Simmonds M
\newblock (2001) Effects of membrane cholesterol on the sensitivity of the
  gabaa receptor to gaba in acutely dissociated rat hippocampal neurones.
\newblock \emph{Neuropharmacology} 40:178 -- 184.

\bibitem{Allen07}
Allen JA, Halverson-Tamboli RA, Rasenick MM
\newblock (2007) Lipid raft microdomains and neurotransmitter signalling.
\newblock \emph{Nat Rev Neurosci} 8:128--140.

\bibitem{lipidSize}
Chiu S, Jakobsson E, Mashl RJ, Scott HL
\newblock (2002) Cholesterol-induced modifications in lipid bilayers: A
  simulation study.
\newblock \emph{Biophysical Journal} 83:1842 -- 1853.

\bibitem{lipidSize2}
Alwarawrah M, Dai J, Huang J
\newblock (2010) A molecular view of the cholesterol condensing effect in dopc
  lipid bilayers.
\newblock \emph{The Journal of Physical Chemistry B} 114:7516--7523
\newblock PMID: 20469902.

\bibitem{Andersen07}
Andersen OS, Roger E.~Koeppe I
\newblock (2007) Bilayer thickness and membrane protein function: An energetic
  perspective.
\newblock \emph{Annual Review of Biophysics and Biomolecular Structure}
  36:107--130
\newblock PMID: 17263662.

\bibitem{Soubias08}
Soubias O, Niu SL, Mitchell DC, Gawrisch K
\newblock (2008) Lipid-rhodopsin hydrophobic mismatch alters rhodopsin helical
  content.
\newblock \emph{J Am Chem Soc} 130:12465--71.

\bibitem{Gruner91}
Gruner SM, Shyamsunder E
\newblock (1991) Is the mechanism of general anesthesia related to lipid
  membrane spontaneous curvature?
\newblock \emph{Ann N Y Acad Sci} 625:685--97.

\bibitem{Levitan14}
Levitan I, Singh DK, Rosenhouse-Dantsker A
\newblock (2014) Cholesterol binding to ion channels.
\newblock \emph{Frontiers in Physiology} 5.

\bibitem{Creighton13}
Thomas CK, Middleton AA
\newblock (2013) Numerically exact correlations and sampling in the
  two-dimensional ising spin glass.
\newblock \emph{Phys. Rev. E} 87:043303.

\bibitem{Kawasaki72}
Kawasaki K
\newblock (1972) \emph{Phase transitions and critical phenomena} eds{} Domb C,
  Green MS
\newblock (Academic Press, Waltham, Massachusetts) Vol.{}~4.

\bibitem{HohenbergHalperin}
Hohenberg P, Halperin B
\newblock (1977) Theory of dynamic critical phenomena.
\newblock \emph{Rev. Mod. Phys.} 49:435--479.

\bibitem{Mortensen07}
Mortensen M, Smart TG
\newblock (2007) Single-channel recording of ligand-gated ion channels.
\newblock \emph{Nat. Protocols} 2:2826--2841.

\bibitem{Wachtel95}
Wachtel RE
\newblock (1995) Relative potencies of volatile anesthetics in altering the
  kinetics of ion channels in bc3h1 cells.
\newblock \emph{Journal of Pharmacology and Experimental Therapeutics}
  274:1355--1361.

\bibitem{Yamashita05}
Yamashita M, Mori T, Nagata K, Yeh J, Narahashi T
\newblock (2005) Isoflurane modulation of neuronal nicotinic acetylcholine
  receptors expressed in human embryonic kidney cells.
\newblock \emph{Anesthesiology} 102:76--84.

\bibitem{MillerCholesterol}
Rankin S, Addona G, Kloczewiak M, Bugge B, Miller K
\newblock (1997) The cholesterol dependence of activation and fast
  desensitization of the nicotinic acetylcholine receptor.
\newblock \emph{Biophysical Journal} 73:2446 -- 2455.

\bibitem{Cardy04}
{Cardy} J
\newblock (2004) {Boundary Conformal Field Theory}.
\newblock \emph{ArXiv High Energy Physics - Theory e-prints}.

\bibitem{Li07}
Li X, Serwanski DR, Miralles CP, Bahr BA, De~Blas AL
\newblock (2007) Two pools of triton x-100-insoluble gabaa receptors are
  present in the brain, one associated to lipid rafts and another one to the
  post-synaptic gabaergic complex.
\newblock \emph{Journal of Neurochemistry} 102:1329--1345.

\bibitem{Zhao13}
Zhao J, Wu J, Veatch SL
\newblock (2013) Adhesion stabilizes robust lipid heterogeneity in
  supercritical membranes at physiological temperature.
\newblock \emph{Biophys J} 104:825--34.

\bibitem{Shachi16}
Katira S, Mandadapu KK, Vaikuntanathan S, Smit B, Chandler D
\newblock (2016) Pre-transition effects mediate forces of assembly between
  transmembrane proteins.
\newblock \emph{eLife} 5:e13150.

\bibitem{HonerkampSmithHydrophobic}
Honerkamp-Smith AR, Machta BB, Keller SL
\newblock (2012) Experimental observations of dynamic critical phenomena in a
  lipid membrane.
\newblock \emph{Phys. Rev. Lett.} 108:265702.

\bibitem{patchClamp}
Hamill O, Marty A, Neher E, Sakmann B, Sigworth F
\newblock (1981) Improved patch-clamp techniques for high-resolution current
  recording from cells and cell-free membrane patches.
\newblock \emph{PflŸegers Archiv} 391:85--100.

\bibitem{Kasianowicz96}
Kasianowicz J, Brandin E, Branton D, Deamer D
\newblock (1996) Characterization of individual polynucleotide molecules using
  a membrane?channel.
\newblock \emph{Proceedings of the National Academy of Sciences}
  93:13770--13773.

\bibitem{Sakmann09}
Sakmann B, Neher E, eds
\newblock (2009) \emph{Single-Channel Recording}
\newblock (Springer Science \& Business Media, Berlin).

\end{thebibliography}

\begin{thebibliography}{1}

\bibitem{Cardy04}
{Cardy} J
\newblock (2004) {Boundary Conformal Field Theory}.
\newblock \emph{ArXiv High Energy Physics - Theory e-prints}.

\bibitem{patchClamp}
Hamill O, Marty A, Neher E, Sakmann B, Sigworth F
\newblock (1981) Improved patch-clamp techniques for high-resolution current
  recording from cells and cell-free membrane patches.
\newblock \emph{PflŸegers Archiv} 391:85--100.

\bibitem{Kasianowicz96}
Kasianowicz J, Brandin E, Branton D, Deamer D
\newblock (1996) Characterization of individual polynucleotide molecules using
  a membrane?channel.
\newblock \emph{Proceedings of the National Academy of Sciences}
  93:13770--13773.

\bibitem{Sakmann09}
Sakmann B, Neher E, eds
\newblock (2009) \emph{Single-Channel Recording}
\newblock (Springer Science \& Business Media, Berlin).

\end{thebibliography}

\end{document}